\documentclass[12pt,titlepage]{article}

\begin{document}

\title{Confinement of fermions by mixed vector-scalar linear potentials in
two-dimensional space-time}
\date{}
\author{Antonio S. de Castro \\
\\
UNESP - Campus de Guaratinguet\'{a}\\
Departamento de F\'{\i}sica e Qu\'{\i}mica\\
Caixa Postal 205\\
12516-410 Guaratinguet\'{a} SP - Brasil\\
\\
Electronic mail: castro@feg.unesp.br}
\maketitle

\begin{abstract}
The problem of confinement of fermions in 1+1 dimensions is approached with
a linear potential in the Dirac equation by considering a mixing of Lorentz
vector and scalar couplings. Analytical bound-states solutions are obtained
when the scalar coupling is of sufficient intensity compared to the vector
coupling.
\end{abstract}

The Coulomb potential of a point electric charge in a 1+1 dimension,
considered as the time component of a Lorentz vector, is linear and so it
provides a constant electric field always pointing to, or from, the point
charge. This problem is related to the confinement of fermions in the
Schwinger and in the massive Schwinger models \cite{col1}-\cite{col2} and in
the Thirring-Schwinger model \cite{fro}. It is frustrating that, due to the
tunneling effect (Klein\'{}s paradox), there are no bound states for this
kind of potential regardless of the strength of the potential \cite{cap}-%
\cite{gal}. The linear potential, considered as a Lorentz scalar, is also
related to the quarkonium model in one-plus-one dimensions \cite{hoo}-\cite
{kog}. Recently it was incorrectly concluded that even in this case there
is just one bound state \cite{bha}.
Later, the proper solutions for this last problem were found \cite{cas}-\cite
{hil}. However, it is well known from the quarkonium phenomenology in the
real 3+1 dimensional world that the best fit for meson spectroscopy is found
for a convenient mixture of vector and scalar potentials put by hand in the
equations (see, \textit{e.g.}, \cite{luc}). Therefore, the problem of
confinement of fermions by a linear potential in 1+1 dimensions deserves a
more general analyses. With this in mind, we approach in the present paper
the Dirac equation in one-plus-one dimensions with a linear potential
considering it as a more general mixing of Lorentz vector and scalar
couplings. It is found that there are analytical bound-state solutions on
condition that the scalar component of the potential is of sufficient
strength compared to the vector component ($|V_{s}|\geq |V_{t}|$). As a
by-product, the present approach also provides the opportunity to find that
there exist relativistic confining potentials providing no bound-state
solutions in the nonrelativistic limit. Although we shall confine our
discussion to the vector-scalar mixing, the inclusion of a pseudoscalar
potential can be allowed. The hearth of the matter is a unitary
transformation similar to that one recently applied to the
vector-pseudoscalar mixing in 3+1 dimensions \cite{vai}. That unitary
transformation leads to a Sturm-Liouville eigenvalue problem for the upper
component of the Dirac spinor.

Let us begin by presenting the Dirac equation in 1+1 dimensions. In the
presence of a time-independent potential the 1+1 dimensional
time-independent Dirac equation for a fermion of rest mass $m$ reads

\begin{equation}
\mathcal{H}\Psi =E\Psi  \label{eq1}
\end{equation}

\begin{equation}
\mathcal{H}=c\alpha p+\beta mc^{2}+\mathcal{V}  \label{eq1a}
\end{equation}

\noindent where $E$ is the energy of the fermion, $c$ is the velocity of
light and $p$ is the momentum operator. $\alpha $ and $\beta $ are Hermitian
square matrices satisfying the relations $\alpha ^{2}=\beta ^{2}=1$, $%
\left\{ \alpha ,\beta \right\} =0$. From the last two relations it steams
that both $\alpha $ and $\beta $ are traceless and have eigenvalues equal to
$-$1, so that one can conclude that $\alpha $ and $\beta $ are
even-dimensional matrices. One can choose the 2$\times $2 Pauli matrices
satisfying the same algebra as $\alpha $ and $\beta $, resulting in a
2-component spinor $\Psi $. The positive definite function $|\Psi |^{2}=\Psi
^{\dagger }\Psi $, satisfying a continuity equation, is interpreted as a
probability position density and its norm is a constant of motion. This
interpretation is completely satisfactory for single-particle states \cite
{tha}. We use $\alpha =\sigma _{1}$ and $\beta =\sigma _{3}$. For the
potential matrix we consider
\begin{equation}
\mathcal{V}=1V_{t}+\beta V_{s}+\alpha V_{e}+\beta \gamma ^{5}V_{p}
\label{eq2}
\end{equation}

\noindent where $1$ stands for the 2$\times $2 identity matrix and $\beta
\gamma ^{5}=\sigma _{2}$. This is the most general combination of Lorentz
structures for the potential matrix because there are only four linearly
independent 2$\times $2 matrices. The subscripts for the terms of potential
denote their properties under a Lorentz transformation: $t$ and $e$ for the
time and space components of the 2-vector potential, $s$ and $p$ for the
scalar and pseudoscalar terms, respectively.

Defining the spinor $\psi $ as
\begin{equation}
\psi =\exp \left( \frac{i}{\hbar }\Lambda \right) \Psi  \label{eq5}
\end{equation}

\noindent where
\begin{equation}
\Lambda (x)=\int^{x}dx^{\prime }\frac{V_{e}(x^{\prime })}{c}  \label{eq6}
\end{equation}

\noindent the space component of the vector potential is gauged away

\begin{equation}
\left( p+\frac{V_{e}}{c}\right) \Psi =\exp \left( \frac{i}{\hbar }\Lambda
\right) p\psi  \label{eq7}
\end{equation}

\noindent so that the time-independent Dirac equation can be rewritten as
follows:

\begin{equation}
H\psi =E\psi  \label{eq8}
\end{equation}

\begin{equation}
H=\sigma _{1}cp+\sigma _{2}V_{p}+\sigma _{3}\left( mc^{2}+V_{s}\right)
+1V_{t}  \label{eq8a}
\end{equation}

\noindent showing that the space component of a vector potential only
contributes to change the spinors by a local phase factor. Introducing the
unitary operator

\begin{equation}
U(\theta )=\exp \left( -i\frac{\theta }{2}\sigma _{1}\right)  \label{eq9}
\end{equation}

\noindent where $\theta $ is a real quantity such that $-\pi \leq \theta
\leq \pi $, the transform of the Hamiltonian (\ref{eq8a}) takes the form

\begin{eqnarray}
\widetilde{H} &=&UHU^{-1}=\sigma _{1}cp+\sigma _{2}\left[ V_{p}\cos \left(
\theta \right) -\left( mc^{2}+V_{s}\right) \sin \left( \theta \right) \right]
\nonumber \\
&&\qquad  \label{eq10} \\
&&\qquad \quad \quad +\sigma _{3}\left[ \left( mc^{2}+V_{s}\right) \cos
\left( \theta \right) -V_{p}\sin \left( \theta \right) \right] +V_{t}
\nonumber
\end{eqnarray}

\noindent In terms of the upper and the lower components of the transform of
the spinor $\psi $ under the action of the operator $U$:
\begin{equation}
\widetilde{\psi }=U\psi =\left(
\begin{array}{c}
\phi \\
\chi
\end{array}
\right)  \label{eq11}
\end{equation}

\noindent and, moreover, choosing

\begin{equation}
V_{t}=V_{s}\cos \left( \theta \right) -V_{p}\sin \left( \theta \right)
\label{eq12}
\end{equation}

\noindent \noindent the Dirac equation decomposes into:

\begin{eqnarray}
&&-\hbar ^{2}c^{2}\phi ^{\prime \prime }+\left\{ \hbar c\left[ V_{s}^{\prime
}\sin \left( \theta \right) +V_{p}^{\prime }\cos \left( \theta \right)
\right] +\left[ \left( mc^{2}+V_{s}\right) \sin \left( \theta \right)
+V_{p}\cos \left( \theta \right) \right] ^{2}\right.  \nonumber \\
&&  \label{13} \\
&&\left. +2\left[ E+mc^{2}\cos \left( \theta \right) \right] \left[
V_{s}\cos \left( \theta \right) -V_{p}\sin \left( \theta \right) \right]
-\left[ E^{2}-m^{2}c^{4}\cos ^{2}\left( \theta \right) \right] \right\} \phi
=0  \nonumber
\end{eqnarray}

\mathstrut

\begin{equation}
\chi =i\frac{-\hbar c\phi ^{\prime }+\left[ \left( mc^{2}+V_{s}\right) \sin
\left( \theta \right) +V_{p}\cos \left( \theta \right) \right] \phi }{%
E+mc^{2}\cos \left( \theta \right) }  \label{eq14}
\end{equation}

\noindent where the prime denotes differentiation with respect to $x$. In
terms of the upper and the lower components the spinor is normalized as

\begin{equation}
\int_{-\infty }^{+\infty }\left( |\phi |^{2}+|\chi |^{2}\right) dx=1
\label{eq14b}
\end{equation}

\noindent so that both $\phi $ and $\chi $ are square integrable functions.

Now, we shall restrict our discussion to linear potentials with $V_{p}=0$,
namely

\begin{eqnarray}
V_{s} &=&a_{s}|x|  \nonumber \\
&&  \label{eq15} \\
V_{t} &=&V_{s}\cos (\theta )  \nonumber
\end{eqnarray}

\noindent \textit{i.e.}, $|V_{s}|\geq |V_{t}|.$ Thus, Eq.(\ref{13}) becomes
the Schr\"{o}dinger-like equation

\smallskip

\begin{eqnarray}
&&-\hbar ^{2}c^{2}\phi ^{\prime \prime }+\left\{ a_{s}^{2}\sin ^{2}(\theta
)x^{2}+2a_{s}\left[ E\cos \left( \theta \right) +mc^{2}\right] |x|\right.
\nonumber \\
&&\qquad  \label{eq16} \\
&&\qquad \quad \quad \left. \qquad +\varepsilon (x)\hbar ca_{s}\sin (\theta
)-\left( E^{2}-m^{2}c^{4}\right) \right\} \phi =0  \nonumber
\end{eqnarray}

\noindent where $\varepsilon (x)=x/|x|$, for $x\neq 0$. It is clear that Eq.(%
\ref{eq16}) allows two distinct classes of solutions depending on $\sin
(\theta )$.

For the class $\sin (\theta )=0$, we define

\begin{equation}
\zeta _{\pm }=\alpha _{\pm }|x|+\beta _{\pm }  \label{eq17}
\end{equation}

\noindent where the plus sign corresponds to $V_{t}=V_{s}$ ($\cos (\theta
)=1 $) and the minus sign corresponds to $V_{t}=-V_{s}$ ($\cos (\theta )=-1$%
). Furthermore,

\begin{eqnarray}
\alpha _{\pm } &=&\pm \left[ \frac{2a_{s}}{\hbar ^{2}c^{2}}\left( E\pm
mc^{2}\right) \right] ^{1/3}  \nonumber \\
&&  \label{eq18} \\
\beta _{\pm } &=&\mp \frac{\alpha _{\pm }}{2a_{s}}\left( E\mp mc^{2}\right)
\nonumber
\end{eqnarray}

\noindent so that Eq.(\ref{eq16}) turns into the Airy differential equation

\begin{equation}
\frac{d^{2}\phi (\zeta _{\pm })}{d\zeta _{\pm }^{2}}-\zeta _{\pm }\phi
(\zeta _{\pm })=0  \label{eq19}
\end{equation}

\noindent which has square integrable solutions expressed in terms of the
Airy functions \cite{abr}: $\phi (\zeta _{\pm })=A_{\pm }\,Ai(\zeta _{\pm })$%
, where $A_{\pm }\mathcal{\ }$is a normalization constant. The joining
condition of $\phi $ and its derivative at $x=0$ leads to the quantization
conditions

\begin{eqnarray}
Ai(\beta _{\pm }) &=&0\;\;\;\;\;{\rm{ for \; odd \; parity \; solutions}}  \nonumber
\\
&&  \label{eq19a} \\
Ai^{\prime }(\beta _{\pm }) &=&0\;\;\;\;\;{\rm{ for \; even \; parity \; solutions}}
\nonumber
\end{eqnarray}

\noindent These quantization conditions have solutions only for $\beta _{\pm
}<0$ and some of them is listed are Table I \cite{abr}. One can see that $%
\beta _{\pm }<0$ always corresponds to $|E|>mc^{2}$. Substitution of the
roots of $Ai(\beta _{\pm })$ and $Ai^{\prime }(\beta _{\pm })$ into (\ref
{eq18}) allow us to obtain the possible energies as the solutions of a
forth-degree algebraic equation:

\begin{equation}
E^{4}\mp 2mc^{2}E^{3}\pm 2m^{3}c^{6}E-\left[ m^{4}c^{8}+\left( 2\hbar
ca_{s}\right) ^{2}|\beta _{\pm }|^{3}\right] =0  \label{eq19b}
\end{equation}

\noindent Instead of giving explicit solutions to this algebraic equation in
terms of radicals, we satisfy ourselves verifying that the roots of Eq. (\ref
{eq19b}) always satisfy the requirement $|E|>mc^{2}$ by using the
Descartes\'{} rule of signs (henceforth DRS). The DRS states that an
algebraic equation with real coefficients $a_{k}\lambda
^{k}+...+a_{1}\lambda +a_{0}=0$ the difference between the number of changes
of signs in the sequence $a_{k},...,a_{1},a_{0}$ and the number of positive
real roots is an even number or zero, with a root of multiplicity $k$
counted as $k$ roots and not counting the null coefficients (see,\textit{\
e.g.}, \cite{sal}-\cite{bar}). The verification of the existence of
solutions for $E>mc^{2}$ is made simpler if we write $E=mc^{2}+\delta $. We
get

\begin{eqnarray}
\delta ^{4}+2mc^{2}\delta ^{3}-\left( 2\hbar ca_{s}\right) ^{2}|\beta
_{+}|^{3} &=&0  \nonumber \\
&&  \label{eq19c} \\
\delta ^{4}+6mc^{2}\delta ^{3}+12m^{2}c^{4}\delta ^{2}+8m^{3}c^{6}\delta
&-&\left( 2\hbar ca_{s}\right) ^{2}|\beta _{-}|^{3} =0  \nonumber
\end{eqnarray}

\noindent Observing the difference of signs among the coefficients of the
leading coefficient and the lowest degree \noindent it becomes clear that
there exist positive roots. In this particular case the DRS assures that
there exists just one solution, since there is only one change of sign in
the sequence of coefficients of (\ref{eq19c}). It is interesting to note
that this result is true whatever the fermion masses and the coupling
constant. The very same conclusion for $E<mc^{2}$ can be obtained by
observing that the upper-sign solutions ($V_{t}=V_{s}$) are mapped into the
lower-sign solutions ($V_{t}=-V_{s}$), and vice-versa, by the change $%
E\rightarrow -E$.

On the other hand, for the class $\sin (\theta )\neq 0$, corresponding to $%
|V_{t}|<|V_{s}|$, we define

\[
\xi =\frac{\xi _{0}}{x_{0}}\left( |x|+x_{0}\right)
\]

\noindent where

\begin{eqnarray}
x_{0} &=&\frac{E\cos \left( \theta \right) +mc^{2}}{a_{s}\sin ^{2}(\theta )}
\nonumber \\
&&  \label{eq20} \\
\xi _{0} &=&\sqrt{\frac{2|a_{s}\sin (\theta )|}{\hbar c}}x_{0}  \nonumber
\end{eqnarray}

\noindent Moreover,

\begin{equation}
\nu =-1+\frac{\left[ E+mc^{2}\cos \left( \theta \right) \right]
^{2}-m^{2}c^{4}\cos \left( 2\theta \right) }{2\hbar ca_{s}\sin ^{3}(\theta )}
\label{eq21}
\end{equation}

\noindent such that

\begin{equation}
-\frac{d^{2}\phi (\xi )}{d\xi ^{2}}+\frac{\xi ^{2}}{4}\phi (\xi )=\left\{
\begin{array}{c}
\left( \nu +1/2\right) \phi (\xi )\;\;\;\;\;x>0 \\
\\
\left( \nu +3/2\right) \phi (\xi )\;\;\;\;\;x<0
\end{array}
\right.  \label{eq21a}
\end{equation}

\noindent whose solutions are the square integrable parabolic cylinder
functions \cite{abr}: $\phi (\xi )=BD_{\nu }(\xi )$, for $x>0$, and $\phi
(\xi )=CD_{\nu +1}(\xi )$, for $x<0$. $B$ and $C$ are normalization
constants. Making use of the recurrence formulas

\begin{eqnarray}
D_{\nu }^{\prime }(z)-\frac{z}{2}D_{\nu }(z)+D_{\nu +1}(z) &=&0  \nonumber \\
&&  \label{eq22} \\
D_{\nu }^{\prime }(z)+\frac{z}{2}D_{\nu }(z)-\nu D_{\nu -1}(z) &=&0
\nonumber
\end{eqnarray}

\noindent and the matching conditions at $x=0$, the quantization condition is

\begin{equation}
D_{\nu +1}\left( \xi _{0}\right) =\pm \sqrt{\nu +1}D_{\nu }\left( \xi
_{0}\right)  \label{eq23}
\end{equation}

\noindent where $\xi _{0}$ is given by (\ref{eq20}). By solving the
quantization condition (\ref{eq23}) for $\nu $ imposing that the solutions
of (\ref{eq21a}) vanish for $\xi \rightarrow +\infty $, one obtains the
possible energy levels by inserting those allowed values of $\nu $ in (\ref
{eq20}):

\[
E=-mc^{2}\cos (\theta )\pm \sqrt{m^{2}c^{4}\cos \left( 2\theta \right)
+2\hbar ca_{s}\left( \nu +1\right) \sin ^{3}(\theta )}
\]

\noindent The numerical computation of (\ref{eq23}) is substantially simpler
when $D_{\nu +1}\left( \xi _{0}\right) $ is written in terms of $D_{\nu
}^{\prime }\left( \xi _{0}\right) $:

\begin{equation}
D_{\nu }^{\prime }\left( \xi _{0}\right) =\left[ \frac{\xi _{0}}{2}\mp \sqrt{%
\nu +1}\right] D_{\nu }\left( \xi _{0}\right)  \label{eq24}
\end{equation}

\noindent Because the normalization of the spinor is not important for the
calculation of the spectrum, one can arbitrarily choose $D_{\nu }\left( \xi
_{0}\right) =1$. By using a fourth-fifth order Runge-Kutta method \cite{fm}
an infinite sequence of allowed values of $\nu $ are found corresponding to
each sign in (\ref{eq24}). The lowest states are given in Table II for $\xi
_{0}=1.$

It is worthwhile to note that the first class of solutions ($|V_{s}|=|V_{t}|$%
) is independent of the sign of $a_{s}$ whereas the second class of
solutions ($|V_{s}|>|V_{t}|$)depends on $|a_{s}\sin (\theta )|$. This
observation permit us to conclude that even a ``repulsive'' potential can be
a confining potential. This peculiar effect is because the scalar potential
behaves like an $x$-dependent rest mass \cite{tha}. Nevertheless, only
potentials with $a_{s}>0$ in the mixing $a_{t}=a_{s}\cos (\theta )$ are
confining potentials in the nonrelativistic approximation and the case $%
V_{t}=-V_{s}$ reduces, for any sign of $a_{s}$, to the case of a free
fermion. It is well known that a confining potential in the nonrelativistic
approach is not confining in the relativistic approach when it is considered
as a Lorentz vector. It surprising that relativistic confining potentials
may result in nonconfinement in the nonrelativistic approach. This last
phenomenon is a consequence of the fact that scalar and vector potentials
couples differently in the Dirac equation whereas there is no such
distinction between scalar and vector potentials in the Schr\"{o}dinger
equation. Therefore the results exposed in this paper, beyond to present a
generalization of previous results, might be of relevance to the quarkonium
phenomenology \textbf{\ }in a four-dimensional space-time.

\smallskip

\bigskip

\bigskip

\noindent \textbf{Acknowledgments}

This work was supported in part through funds provided by CNPq and FAPESP.

\newpage
\begin{table}[tbp]
\caption{The first roots of the Airy function and its first derivative}
\label{t1}
.
\par
\begin{center}
\begin{tabular}{|c|c|}
\hline\hline
$Ai^{\prime }(-|\beta _{\pm }|)=0$ & $Ai(-|\beta _{\pm }|)=0$ \\ \hline
1.019 & 2.338 \\
3.248 & 4.088 \\
4.820 & 5.521 \\ \hline\hline
\end{tabular}
\end{center}
\end{table}

\begin{table}[tbp]
\caption{The lowest solutions of Eq. (28) for $\xi_{0}=1$.}
\label{t2}
\begin{center}
\begin{tabular}{|c|c|}
\hline\hline
$\nu$ \textrm{for the minus sign} & $\nu$ \textrm{for the plus sign} \\
\hline
1.580$\times 10^{-4}$ & 1.525 \\
2.681 & 3.915 \\
5.038 & 6.210 \\ \hline\hline
\end{tabular}
\end{center}
\end{table}

\end{document}